\documentclass[aps,prc,twocolumn, floatfix]{revtex4}
\usepackage{graphicx}
\usepackage{dcolumn}
\usepackage{bm}

\begin{document}

\title{Effects of curvature of the symmetry energy in Sn+Sn reactions at 270 MeV/nucleon}

\author{Ya-Fei Guo$^{1,2}$}
\author{Gao-Chan Yong$^{1,2}$}
\email[]{yonggaochan@impcas.ac.cn}

\affiliation{$^1$Institute of Modern Physics, Chinese Academy of Sciences, Lanzhou 730000, China\\
$^2$School of Nuclear Science and Technology, University of Chinese Academy of Sciences, Beijing 100049, China}

\begin{abstract}

Based on the Isospin-dependent Boltzmann-Uehling-Uhlenbeck (IBUU) transport model, the isotope Sn+Sn reactions at 270 MeV/nucleon are investigated. It is shown that nucleon and meson observables in the Sn+Sn reactions at 270 MeV/nucleon cannot effectively probe the high-density symmetry energy. These observables, however, are sensitive to the curvature of the symmetry energy. It thus sheds light on the study of the high-density behavior of the nuclear symmetry energy by extracting the curvature of the symmetry energy through the interpretation of forthcoming RIBF/RIKEN related data.

\end{abstract}

\maketitle


As saturation of the strong interaction, nucleons usually are at around saturation or low densities. In some special astrophysical environment, such as in neutron stars \cite{1Lat,2Lat}, they can be at supra-normal densities. In terrestrial experiments, nucleons in nuclei can be compressed to form supra-normal density nuclear matter via heavy-ion collisions \cite{3Li}. To describe density and isospin-dependence of nuclear matter, the equation of state (EoS) of asymmetric nuclear matter is frequently mentioned in the literature and usually expressed at density $\rho$ as
\begin{equation}
E(\rho,\delta)=E(\rho, 0)+E_{sym}(\rho)\delta^2+O(\delta^4),
\end{equation}
where $\delta=(\rho_n-\rho_p)/\rho$ is the neutron-proton asymmetry and $E_{sym}(\rho)$ is the density-dependent nuclear symmetry energy.
The density-dependent nuclear symmetry energy can then be Taylor expanded around saturation density $\rho_0$ as \cite{9vid}
\begin{equation}\label{esym2}
E_{sym}(\rho)=E_{sym}(\rho_0)+L\frac{\rho-\rho_0}{3\rho_0}+
\frac{K_{sym}}{2}\left(\frac{\rho-\rho_0}{3\rho_0}\right)^2 +...,
\end{equation}
where $L$ and $K_{sym}$ are slope and curvature parameters of the symmetry energy at $\rho_0$,
\begin{equation}
L=3\rho_0\frac{\partial E_{sym}(\rho)}{\partial\rho}\bigg|_{\rho=\rho_0}, K_{sym}=9\rho_0^2\frac{\partial^2 E_{sym}(\rho)}{\partial\rho^2}\bigg|_{\rho=\rho_0}.
\end{equation}
The values $E_{sym}(\rho_0)$, $L$ and $K_{sym}$ of the symmetry energy at $\rho_0$ have been roughly constrained from analysis of terrestrial nuclear laboratory experiments and astrophysical observations over the last two decades with $E_{sym}(\rho_0)=31.7\pm3.2$ MeV, $L=58.7\pm28.1$ MeV, $-400 MeV<K_{sym}<100$ MeV \cite{10Oer,11Li,12Te,13Zhang}.

The high-density symmetry energy is controversial \cite{Guo14}, but closely related to a series of properties of neutron stars, e.g., their radii and cooling rates \cite{1Lat,2Lat,Vil04,Ste05}, the gravitational-wave frequency \cite{gwf,gwf2}, the gamma-ray bursts \cite{grb}, the neutron-star tidal deformability \cite{tsang19} as well as the r-process nucleosynthesis \cite{rpn1,rpn2,rpn3} in neutron star mergers \cite{GWth,GW170817}. The larger uncertain of the high-density symmetry energy is closely related to the less constraints of the $K_{sym}$. In this study, we show how to get information of the $K_{sym}$ from the ongoing RIBF/RIKEN related data. Although the isotope Sn+Sn reaction at 270 MeV/nucleon at RIBF/RIKEN in Japan cannot effectively probe the high-density symmetry energy \cite{yong20191}, the extraction of the curvature of the symmetry energy through interpretation of related data can give implication on the high-density behavior of the symmetry energy Since the slope $L$ of the symmetry energy at saturation density has been constrained with high probability \cite{10Oer}.


The used Isospin-dependent Boltzmann-Uehling-Uhlenbeck (IBUU) transport model \cite{16Li} has appropriately taken the effects of the short-range correlations into account \cite{17yong}, with neutron and proton density distributions in initial colliding nuclei given by the Skyrme-Hartree-Fock with Skyrme M* force parameters \cite{18Fr} and their initial momentum distributions given by the parameterized high-momentum tails \cite{17yong,19Su,20He}.
Details on the baryon-baryon scattering cross section and pion production can be found in Ref.~\cite{21yong}. In the mean-field approximation, the isospin and momentum-dependent single-nucleon potential reads
\begin{eqnarray}
U(\rho,\delta,\vec{p},\tau)&=&A_{u}(x)\frac{\rho_{\tau^\prime}}{\rho_0}+A_{l}(x)\frac{\rho_\tau}{\rho_0}+
B(\frac{\rho}{\rho_0})^\sigma(1-x\delta^2)\nonumber\\
&&-8x\tau\frac{B}{\sigma+1}\frac{\rho^{\sigma-1}}{{\rho_0}^\sigma}\delta \rho_{\tau^\prime}\nonumber\\
&&+\frac{2C_{\tau,\tau}}{\rho_0}\int d^3p^\prime\frac{f_\tau(\vec{r},\vec{p}^\prime)}{1+(\vec{p}-\vec{p}^\prime)^2/\Lambda^2}\nonumber\\
&&+\frac{2C_{\tau,\tau'}}{\rho_0}\int d^3p^\prime\frac{f_{\tau^\prime}(\vec{r},\vec{p}^\prime)}{1+(\vec{p}-\vec{p}^\prime)^2/\Lambda^2},
\label{buupotential}
\end{eqnarray}
where $\tau$,$\tau^\prime=1/2(-1/2)$ is for neutron (proton). $\rho_0$ is the nuclear saturation density. $\delta=(\rho_n-\rho_p)/(\rho_n+\rho_p)$ is the isospin asymmetry with $\rho_n$ and $\rho_p$ being the neutron and proton local densities. The parameters $A_u(x)=33.037-125.34x$ MeV, $A_l(x)=-166.963+125.34x$ MeV, $B=141.96$ MeV, $C_{\tau,\tau}=18.177$ MeV, $C_{\tau,\tau^\prime}=-178.365$ MeV, $\sigma=1.265$, and $\Lambda=630.24$ MeV/c \cite{21yong}.
In Eq.~(\ref{buupotential}), one can keep the parameters $E_{sym}(\rho_0)$ and slope $L$ fixed but varying the $K_{sym}$ if a density-dependent $x$ parameter in Eq.~(\ref{buupotential}) is used \cite{22yong}. To probe the curvature of the symmetry energy, we in this study fix $E_{sym}(\rho_0)=30$ MeV and $L=40$ MeV with density-dependent $x$ parameter for $K_{sym}=0$ MeV
\begin{eqnarray}\label{k0}
x&=&0.5902(\rho/\rho_0)^6-5.57(\rho/\rho_0)^5+21.146(\rho/\rho_0)^4-\nonumber\\
&&41.405(\rho/\rho_0)^3+44.668(\rho/\rho_0)^2-26.429(\rho/\rho_0)\nonumber\\
&&+7.9706
\end{eqnarray}
and for $K_{sym}=-400$ MeV,
\begin{eqnarray}\label{km400}
x&=&0.799(\rho/\rho_0)^6-7.3573(\rho/\rho_0)^5+26.432(\rho/\rho_0)^4-\nonumber\\
&&46.748(\rho/\rho_0)^3
+42.138(\rho/\rho_0)^2-17.661(\rho/\rho_0)\nonumber\\
&&+3.4464.
\end{eqnarray}
A slope of $L=80$ MeV is also used as a variation in later discussions.
\begin{figure}[tbh]
\begin{center}
\includegraphics[width=0.5\textwidth]{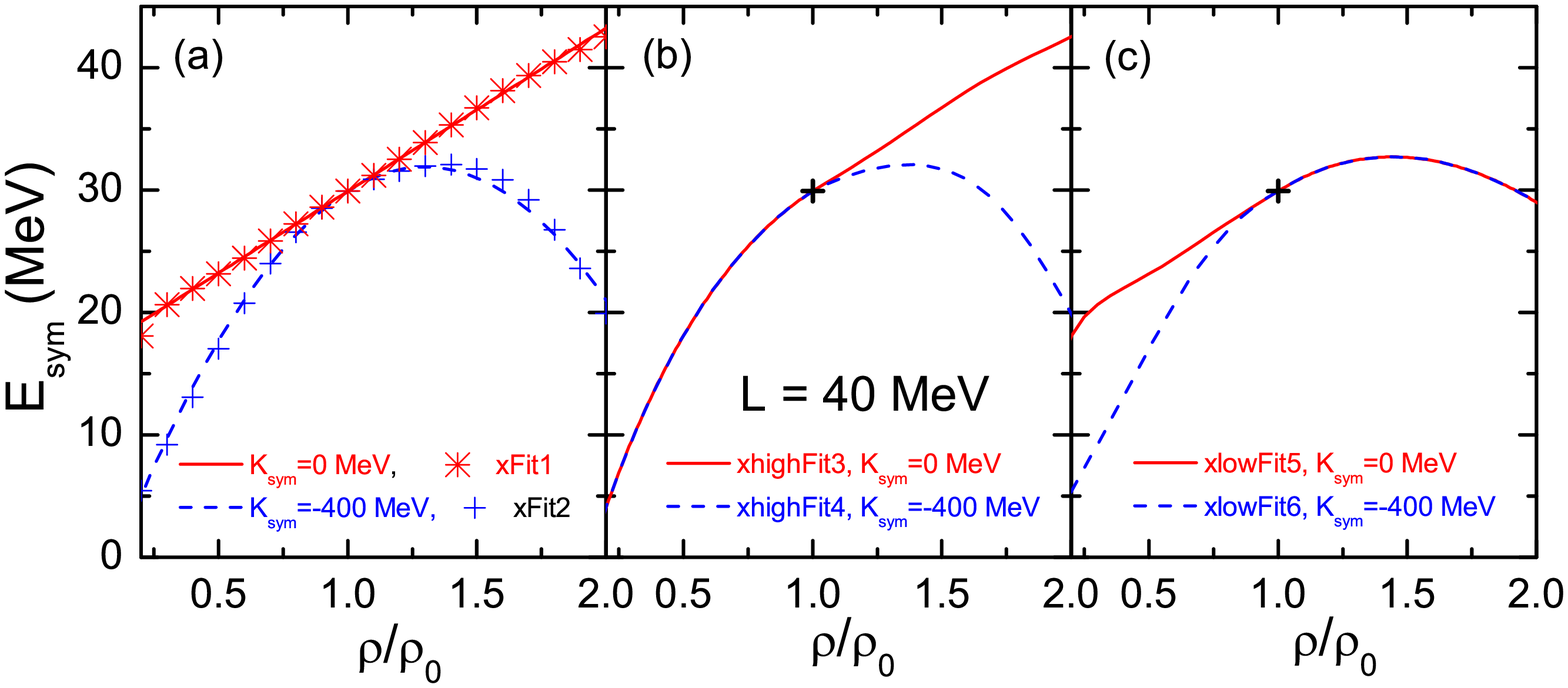}
\end{center}
\caption{Density-dependent symmetry energies with different curvatures of the symmetry energy at low and high densities. In the left panel, the lines show the density-dependent symmetry energies with $E_{sym}(\rho_0)=30$ MeV and $L=40$ MeV fixed but varying the curvature $K_{sym}$ obtained by Eq.~(\ref{esym2}) and the symbols ``xFit1'' and ``xFit2'' denote the symmetry energies with the same $E_{sym}(\rho_0)$, $L$ and $K_{sym}$ derived from Eq.~(\ref{buupotential}) with $x$ from Eq.~(\ref{k0}) and Eq.~(\ref{km400}), respectively. In the middle panel, the low-density ($0-\rho_{0}$) symmetry energy is derived from Eq.~(\ref{buupotential}) with $x=0.947$ (corresponding the same $L$) and the high-density symmetry energies are taken from symmetry energies of ``xFit1'' and ``xFit2'', the symmetry energies in the whole density range are re-labeled by ``xhighFit3'' and ``xhighFit4'', respectively. In the right panel, the high-density symmetry energy is derived from Eq.~(\ref{buupotential}) with $x=0.947$ and the low-density parts are taken from the symmetry energies of ``xFit1'' and ``xFit2'', the symmetry energies in the whole density range are re-labeled by ``xlowFit5'' and ``xlowFit6'', respectively.}\label{esymk}
\end{figure}
Fig.~\ref{esymk} shows 6 density-dependent symmetry energies with the same $E_{sym}(\rho_0)$ and $L$ settings but different curvature $K_{sym}$. In panel (a), whether below or above $\rho_{0}$, the same $K_{sym}$ of the symmetry energy is employed. While in panels (b) or (c), the low or high-density parts of the symmetry energy is fixed (both $x=0.947$ in Eq.~(\ref{buupotential})), but for the left parts the curvature $K_{sym}$ at $\rho_{0}$ is varied. The purpose of these settings on the symmetry energy in Fig.~\ref{esymk} is to see if some observables in Sn+Sn reactions at 270 MeV/nucleon are sensitive to the $K_{sym}$ and to see which part of the symmetry energy (low or high-density symmetry energy) affects these observables.


The radioactive beam facilities worldwide provide a great opportunity to study the properties of isospin asymmetric nuclear matter. At RIBF/RIKEN in Japan, the radioactive beams of isotope Sn +Sn reactions at intermediate energies are currently used to study the density dependence of the nuclear symmetry energy \cite{14sy,15sh}. Specifically, the neutron-rich isotope $^{132}$Sn+$^{124}$Sn and neutron-deficient isotope $^{108}$Sn+$^{112}$Sn reactions are being carried out to probe the density-dependent symmetry energy.

\begin{figure}[tbh]
\begin{center}
\includegraphics[width=0.5\textwidth]{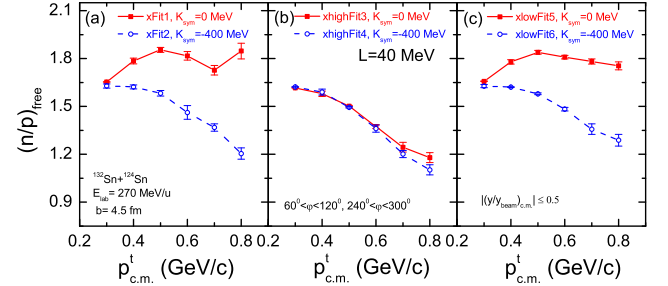}
\end{center}
\caption{The n/p ratio of squeezed-out free nucleons as a function of transverse momentum $p_t$ in neutron-rich isotope $^{132}$Sn+$^{124}$Sn semi-central reactions at 270 MeV/nucleon with impact parameter b= 4.5 fm, azimuthal angle cuts $60^{\circ}\leq\phi\leq120^{\circ}$ and $240^{\circ}\leq\phi\leq300^{\circ}$ and rapidity cuts $|(y/y_{beam})_{c.m.}|\leq0.5$.}\label{rnp-rich}
\end{figure}
It is generally considered that the squeezed-out particles, which are directly from the participant region thus carry more information about dense matter, are more sensitive to the properties of dense matter \cite{24yong}. With 6 kinds of density-dependent symmetry energies shown in Fig.~\ref{esymk}, free n/p ratios of squeezed-out nucleons as a function of transverse momentum $p_t$ in the semi-central $^{132}$Sn+$^{124}$Sn reactions at 270 MeV/nucleon are demonstrated in Fig.~\ref{rnp-rich}. From the left panel of Fig.~\ref{rnp-rich}, it is clearly seen that free n/p ratio of the squeezed-out nucleons is very sensitive the curvature of the symmetry energy, especially at high transverse momenta. Compared squeezed-out n/p ratio in the middle panel to the right panel, one can see that the sensitivity of the squeezed-out n/p ratio to the symmetry energy is from the low-density part, i.e., the squeezed-out n/p ratio in the semi-central $^{132}$Sn+$^{124}$Sn reaction at 270 MeV/nucleon mainly probes the low-density symmetry energy. However, from Eq.~(\ref{esym2}), one can deduce the high-density behavior of the symmetry energy if one knows $L$ and $K_{sym}$ of the symmetry energy at $\rho_0$. Given current constraints of the value of the slope of nuclear symmetry energy $L$ \cite{10Oer,11Li}, one can readily deduce the high-density behavior of the symmetry energy only if the curvature $K_{sym}$ is constraints. So it is meaningful to extract the information of $K_{sym}$ from the squeezed-out n/p ratio in the semi-central $^{132}$Sn+$^{124}$Sn reactions at 270 MeV/nucleon as guiding shown in the left panel of Fig.~\ref{rnp-rich}.

\begin{figure}[tbh]
\begin{center}
\includegraphics[width=0.5\textwidth]{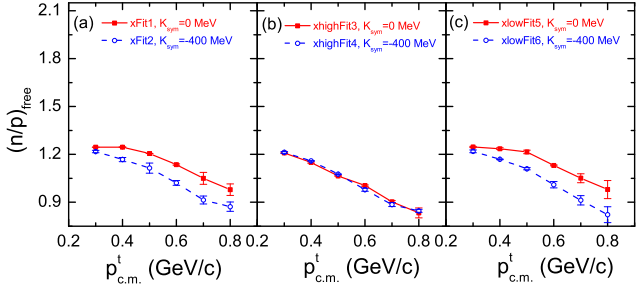}
\end{center}
\caption{Same as Fig.~\ref{rnp-rich}, but for isotope $^{108}$Sn+$^{112}$Sn reactions.}\label{rnp-poor}
\end{figure}
\begin{figure}[tbh]
\begin{center}
\includegraphics[width=0.5\textwidth]{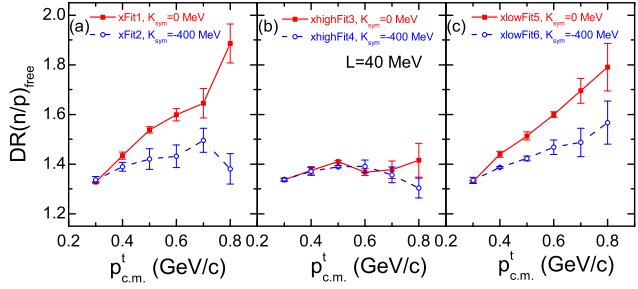}
\end{center}
\caption{Double ratio of the squeezed-out n/p nucleons as a function of transverse momentum $p_t$ in $^{132}$Sn+$^{124}$Sn and $^{108}$Sn+$^{112}$Sn reactions at 270 MeV/nucleon.}\label{double}
\end{figure}
To reduce systematic errors, the neutron-deficient reaction system of the same element is often used as counterpart. Fig.~\ref{rnp-poor} shows the same contents as that demonstrated in Fig.~\ref{rnp-rich}. It is seen from Fig.~\ref{rnp-poor}, the sensitivity of the squeezed-out n/p ratio to the curvature $K_{sym}$ is reduced to some extend. Again, it is shown that the squeezed-out n/p ratio is affected by the low-density symmetry energy. Compared the results shown in Fig.~\ref{rnp-poor} to that shown in Fig.~\ref{rnp-rich}, the results are qualitatively consistent but the whole results in neutron-deficient reactions are lower than in neutron-rich reactions. The double n/p ratio of the squeezed-out n/p nucleons in neutron-rich $^{132}$Sn+$^{124}$Sn and neutron-deficient $^{108}$Sn+$^{112}$Sn reactions is defined as
\begin{equation}
DR(n/p)=\frac{(n/p)_{^{132}Sn+^{124}Sn}}{(n/p)_{^{108}Sn+^{112}Sn}}.
\end{equation}
From the left panel of Fig.~\ref{double}, it is seen that sensitivity of the double n/p ratio of the squeezed-out n/p nucleons to the curvature $K_{sym}$ is kept.

\begin{figure}[tbh]
\begin{center}
\includegraphics[width=0.5\textwidth]{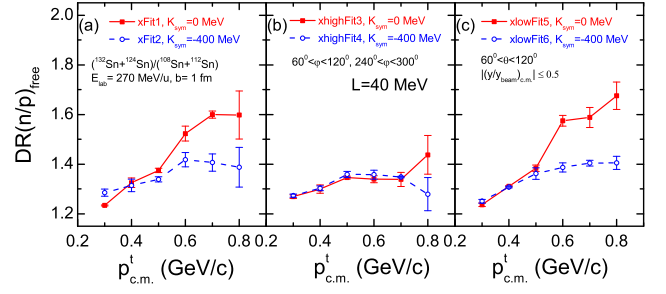}
\end{center}
\caption{Same as Fig.~\ref{double}, but for central collisions.}\label{doubleb1}
\end{figure}
To see the impact parameter dependence of the above discussions, we provide the case with small impact parameter. Fig.~\ref{doubleb1} shows the double ratio of the squeezed-out n/p nucleons as a function of transverse momentum $p_t$ in central $^{132}$Sn+$^{124}$Sn and $^{108}$Sn+$^{112}$Sn reactions at 270 MeV/nucleon. It is seen that the physical results are independence of impact parameter, but the specific values of the double ratio of the squeezed-out n/p nucleons are varied.

\begin{figure}[tbh]
\begin{center}
\includegraphics[width=0.5\textwidth]{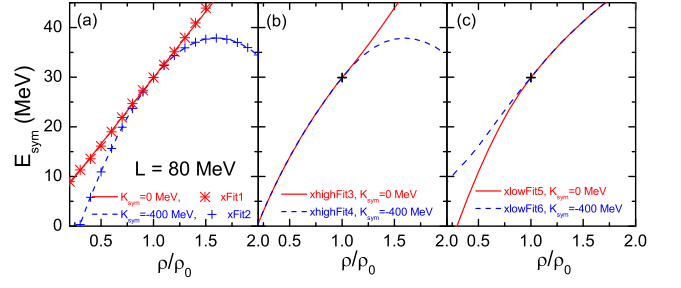}
\end{center}
\caption{Same as Fig.~\ref{esymk}, but for $L$= 80 MeV. Note that $x$= 0.145 is used in Eq.~(\ref{buupotential}) to correspond $L$= 80 MeV}\label{esymk80}
\end{figure}
\begin{figure}[tbh]
\begin{center}
\includegraphics[width=0.5\textwidth]{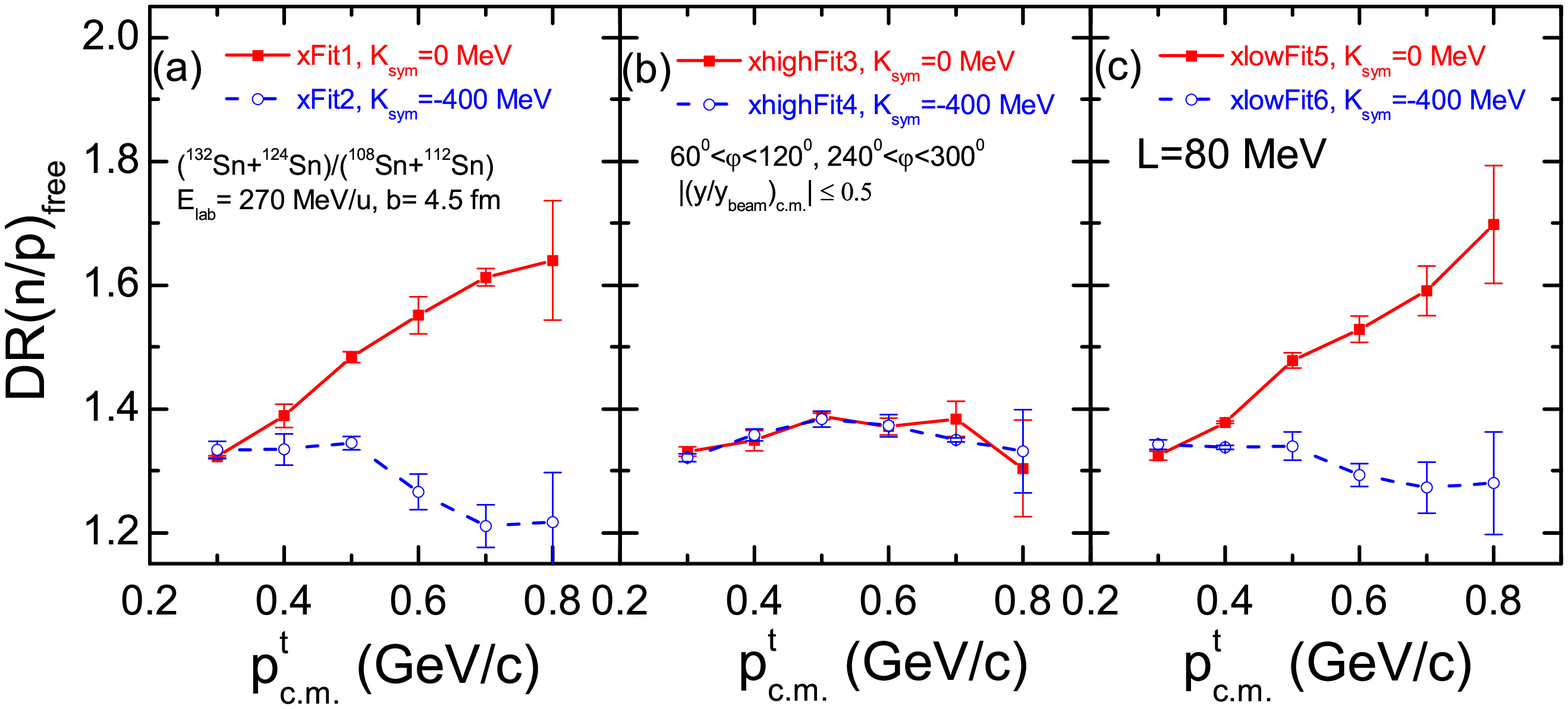}
\end{center}
\caption{Same as Fig.~\ref{double}, but for $L$= 80 MeV.}\label{doubleb2}
\end{figure}
Although the slope $L$ of the symmetry energy at $\rho_{0}$ has been constrained with high probability \cite{10Oer,11Li}, there is still room of uncertainty.
It is thus necessary to see the effects of $L$ on the double ratio of the squeezed-out n/p nucleons in Sn+Sn reactions at 270 MeV/nucleon. To investigate the effects of slope $L$ of the symmetry energy, the slope parameter $L=40$ MeV is replaced by $L=80$ MeV and other conditions are kept. Thus, in Eq.~(\ref{buupotential}) the density-dependent $x$ parameters for $K_{sym}=0$ MeV case reads
\begin{eqnarray}
x&=&0.6924(\rho/\rho_0)^6-6.6431(\rho/\rho_0)^5+25.082(\rho/\rho_0)^4\nonumber\\
&&-47.349(\rho/\rho_0)^3+
47.049(\rho/\rho_0)^2-23.939(\rho/\rho_0)\nonumber\\
&&+5.3267
\end{eqnarray}
and for $K_{sym}=-400$ MeV case
\begin{eqnarray}
x&=&0.1579(\rho/\rho_0)^5-1.239(\rho/\rho_0)^4+\nonumber\\
&&3.8325(\rho/\rho_0)^3-5.9913(\rho/\rho_0)^2\nonumber\\
&&+5.4201(\rho/\rho_0)-2.0336
\end{eqnarray}
and $x$= 0.145 in Eq.~(\ref{buupotential}) for the slope $L=80$ MeV at saturation density is employed. We first provide Fig.~\ref{esymk80}, which is the same as Fig.~\ref{esymk} but with all the density-dependent symmetry energies possess the same slope of $L$= 80 MeV at saturation density. Fig.~\ref{doubleb2} demonstrates the same contents as those shown in Fig.~\ref{double} or Fig.~\ref{doubleb1}. The same physical results are seen as that shown in Fig.~\ref{double} or Fig.~\ref{doubleb1}. Compared the results shown in Fig.~\ref{doubleb2} with Fig.~\ref{double}, the slope $L$ of the symmetry energy really affects the value of the double ratio of the squeezed-out n/p nucleons. And the effects of the curvature $K_{sym}$ on the squeezed-out n/p nucleons also become larger as the increase of the slope of $L$.

\begin{figure}[tbh]
\begin{center}
\includegraphics[width=0.5\textwidth]{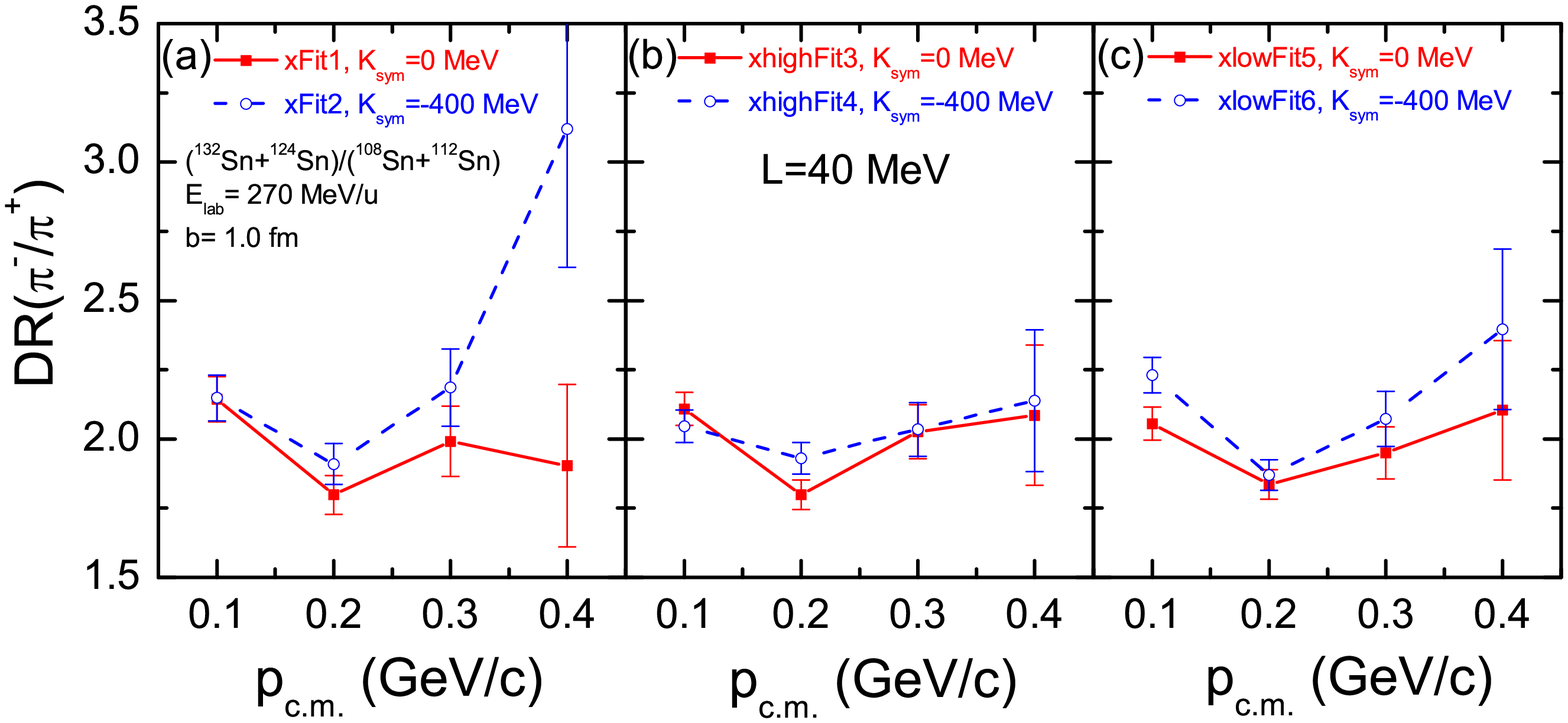}
\end{center}
\caption{Double $\pi^-/\pi^+$ ratio as a function of momentum in $^{132}$Sn+$^{124}$Sn and $^{108}$Sn+$^{112}$Sn central reactions at 270 MeV/nucleon.}\label{drpion}
\end{figure}
Similar to the double ratio of the squeezed-out n/p nucleons as shown in Fig.~\ref{double}, we also plot double $\pi^-/\pi^+$ ratio in the $^{132}$Sn+$^{124}$Sn and $^{108}$Sn+$^{112}$Sn reactions, which reads as
\begin{equation}
DR(\pi^-/\pi^+)=\frac{(\pi^-/\pi^+)_{^{132}Sn+^{124}Sn}}{(\pi^-/\pi^+)_{^{108}Sn+^{112}Sn}}.
\end{equation}
Fig.~\ref{drpion} demonstrates that the double $\pi^-/\pi^+$ ratio can also be used to probe the curvature $K_{sym}$ of the symmetry energy. From panels (b) and (c) of Fig.~\ref{drpion}, one can see that the double $\pi^-/\pi^+$ ratio is somewhat more sensitive to the low-density symmetry energy, which is consistent with that shown in Ref.~\cite{yong20191}. However, one can deduce the high-density behavior of the symmetry energy once the $K_{sym}$ is probed and constrained from RIBF/RIKEN experiments.


In the Sn+Sn reactions at 270 MeV/nucleon, one cannot effectively extract the information on high-density symmetry energy from both nucleon and meson observables. However, the curvature information of the symmetry energy may be obtained from these observables assuming the slope of the symmetry energy is known. Therefore the Sn+Sn reactions at 270 MeV/nucleon, which are being carried out at RIKEN in Japan, are still meaningful on the exploration of the high-density behavior of the nuclear symmetry energy.


This work is supported in part by the National Natural Science Foundation of China under Grant Nos. 11775275, 11435014.


\begin{thebibliography}{00}

\bibitem{1Lat}J. M. Lattimer and M. Prakash, Ap. J. {\bf 550}, 426 (2001).
\bibitem{2Lat}J. M. Lattimer and M. Prakash, Science {\bf 304}, 536 (2004).
\bibitem{3Li}B. A. Li, L. W. Chen, and C. M. Ko, Phys. Rep. {\bf 464}, 113 (2008).
\bibitem{9vid}I. Vida$\tilde{n}$a, C. Provid$\hat{e}$ncia, A. Polls, and A. Rios, Phys. Rev. C {\bf 80}, 045806 (2009).
\bibitem{10Oer}M. Oertel, M. Hempel, T. Kl$\ddot{a}$hn, and S. Typel, Rev. Mod. Phys. {\bf 89}, 015007 (2017).
\bibitem{11Li}B. A. Li and X. Han, Phys. Lett. B {\bf 727}, 276 (2013).
\bibitem{12Te}I. Tews, J. M. Lattimer, A. Ohnishi, and E. E. Kolomeitsev, Ap. J. {\bf 848}, 105 (2017).
\bibitem{13Zhang}N. B. Zhang, B. J. Cai, B. A. Li,W. G. Newton, and J. Xu, Nucl. Sci. Tech. {\bf 28}, 181 (2017).
\bibitem{Guo14}W. M. Guo, G. C. Yong, Y. J. Wang, Q. F. Li, H. F. Zhang, W. Zuo, Phys. Lett. B {\bf 738}, 397 (2014).
\bibitem{Vil04}Adam R. Villarreal and Tod E. Strohmayer, Ap. J. {\bf 614}, L121 (2004).
\bibitem{Ste05}A. W. Steiner, M. Prakash, J.M. Lattimer, P.J. Ellis, Phys. Rep. {\bf 411}, 325 (2005).
\bibitem{gwf}A. Maselli, L. Gualtieri, V. Ferrari, Phys. Rev. D {\bf 88}, 104040 (2013).
\bibitem{gwf2}A. Bauswein, N. Stergioulas, H. T. Janka, Phys. Rev. D {\bf 90}, 023002 (2014).
\bibitem{grb}P. D. Lasky, B. Haskell, V. Ravi, E. J. Howell, D. M. Coward, Phys. Rev. D {\bf 89}, 047302 (2014).
\bibitem{tsang19}C. Y. Tsang, M. B. Tsang, P. Danielewicz, F. J. Fattoyev, W. G. Lynch, Phys. Lett. B {\bf 796}, 1 (2019).
\bibitem{rpn1}S. Goriely, A. Bauswein, H. T. Janka, Ap. J. {\bf 738}, 32 (2011).
\bibitem{rpn2}A. Bauswein, S. Goriely, H. T. Janka, Ap. J. {\bf 773}, 78 (2013).
\bibitem{rpn3}S. Wanajo, Y. Sekiguchi, N. Nishimura, K. Kiuchi, K. Kyutoku, M. Shibata, Ap. J. {\bf 789}, 39 (2014).
\bibitem{GWth}S. Rosswog, International Journal of Modern Physics D {\bf 24}, 1530012 (2015).
\bibitem{GW170817}B. P. Abbott \emph{et al}. (LIGO Scientific Collaboration and Virgo Collaboration), Phys. Rev. Lett. {\bf 119}, 161101 (2017).
\bibitem{yong20191}G. C. Yong, Y. Gao, G. F. Wei, Y. F. Guo, W. Zuo, J. Phys. G: Nucl. Part. Phys. {\bf 46}, 105105 (2019).
\bibitem{16Li}B. A. Li, G. C. Yong, and W. Zuo, Phys. Rev. C {\bf 71}, 014608 (2005).
\bibitem{17yong}G. C. Yong and B. A. Li, Phys. Rev. C {\bf 96}, 064614 (2017).
\bibitem{18Fr}J. Friedrich and P. G. Reinhard, Phys. Rev. C {\bf 33}, 335 (1986).
\bibitem{19Su}R. Subedi et al. (Hall A. Collaboration), Science {\bf 320}, 1476 (2008).
\bibitem{20He}O. Hen et al. (The CLAS Collaboration), Science {\bf 346}, 614 (2014).
\bibitem{21yong}G. C. Yong, Phys. Rev. C {\bf 96}, 044605 (2017).
\bibitem{22yong}Y. F. Guo, G. C. Yong, Phys. Rev. C {\bf 100}, 014617 (2019).
\bibitem{14sy}Symmetry Energy Project, https://groups.nscl.msu.edu/hira/sepweb/pages/home.html.
\bibitem{15sh}R. Shane et al., Nuclear Instruments and Method A {\bf 784}, 513 (2015).
\bibitem{24yong}G. C. Yong, B. A. Li, L. W. Chen, Phys. Lett. B {\bf 650} 344 (2007).

\end{thebibliography}
\end{document}